\documentstyle[twocolumn,twoside,prl,floats,aps]{revtex}
\tighten
\begin{document}
\draft

\vskip 0.4cm

\title{ On the distribution of transmission
 eigenvalues in disordered wires}

\author{M. Caselle}

\address{  Dipartimento di Fisica
Teorica dell'Universit\`a di Torino and
 I.N.F.N., Sezione di Torino\\
 via P.Giuria 1, I-10125 Turin,Italy\\
\parbox{14 cm}{\medskip\rm\indent
We solve the Dorokhov-Mello-Pereyra-Kumar equation which
describes the evolution of an ensamble of disordered wires of increasing
length in  the three cases $\beta=1,2,4$. The solution is
obtained by mapping the problem in that of a suitable
Calogero-Sutherland model. In the $\beta=2$ case our solution is in
complete agreement with that recently found by Beenakker and Rejaei.}}
\maketitle
\narrowtext

During last years an increasing interest has been attracted by the
physics of quantum electronic transport in mesoscopic
systems~\cite{alw}. One of the main reasons for this interest lies in the
high degree of universality of some experimental observations.
In particular it is well known that universal conductance fluctuations
(UCF) occur in small metallic samples at low temperatures.
This universality suggests that UCF could be
described by some relatively simple Hamiltonian, independent of the
particular model or disorder realization and led to the construction of
a Random Matrix Theory (RMT) of quantum transport~\cite{im,mps}, in analogy to
the
Wigner-Dyson RMT for nuclear energy levels.
The central object in this approach is the $N\times N$
transmission matrix $t$ of the conductor (where $N$ is the number of
scattering channels at the Fermi level). All the interesting physical
properties
 can be evaluated once the transmission eigenvalues $T_1,T_2 \cdots
T_N$ of the product $tt^\dagger$ are known. For instance the conductance
$G$ is given by $G=G_0~{\rm Tr}(tt^\dagger)=G_0\sum_n T_n$
(with $G_0=2e^2/h$). A suitable parametrization of the eigenvalues
$T_i$ is given in terms of the Transfer Matrix $M$, whose eigenvalues
$\lambda_i$ are related to the $T_i$ by
$\lambda_{i}\equiv (1-T_{i})/T_{i}$.
The goal is then to find the probability distribution
 $P(\{\lambda_i\})$ for the eigenvalues $\lambda_i$. This is in general
 a difficult problem, but it simplifies in
 the quasi-one-dimensional regime (namely when the length $L$ of the
conductor is much larger than its width $W$). In this case
it can be shown that eigenvalue distribution $P(\{\lambda_i\})$ as a
function of the length $L$ (measured in units of the mean free path
$l$: $s=L/l$)  must
obey the  Dorokhov-Mello-Pereyra-Kumar (DMPK) equation
\begin{equation}
\frac{\partial P}{\partial s}~=
\frac{2}{\gamma}\sum_{i=1}^{N}
\frac{\partial}{\partial\lambda_{i}}\lambda_{i}(1+\lambda_{i})
J\frac{\partial}{\partial\lambda_{i}}J^{-1}~P
\equiv~D~P,
\label{DMPK}
\end{equation}
where $\gamma=\beta N+2-\beta$ and
$\beta\in\{1,2,4\}$ is the symmetry index of the ensemble of scattering
matrices. $\beta=1$ if there is time-reversal symmetry;
$\beta=2$ if the time-reversal symmetry is broken
and $\beta=4$ if time-reversal is conserved, but spin-orbit interactions
are important and the spin-rotation symmetry is broken.
 $J(\{\lambda_{n}\})$ denotes the Jacobian from the matrix to the
eigenvalue space:
\begin{equation}
J(\{\lambda_{n}\})=\prod_{i<j}|\lambda_{j}-\lambda_{i}|^{\beta}.
\label{jacobian}
\end{equation}
 Eq.\ (\ref{DMPK}) was derived
by Dorokhov~\cite{dor} from microscopic considerations (for $\beta=2$)
 and by Mello, Pereyra, and Kumar~\cite{mpk}
(for $\beta=1$) by looking at
the infinitesimal transfer matrix describing the addition
of a thin slice to the wire. Their results were then
generalized to $\beta=2,4$ in Refs.~\cite{ms,cm1}.

At a first glance, solving  eq.(\ref{DMPK}) seems to be an impossible
task (except, obviously, the case in which $N=1$) due to the
interaction among eigenvalues.
However, recently, in two remarkable papers~\cite{br},
Beenakker and Rejaei succeeded
in solving it, in the particular case $\beta=2$, by using Sutherland's
transformation and mapping eq.(\ref{DMPK}) into the Schr\"odinger
equation (in imaginary time) of a set on $N$ non-interacting fermions.

Let us briefly review their result, which will be useful in our
following discussion.
The first step suggested in~\cite{br}
is to choose a new set of variables $\{x_n\}$, related to
the eigenvalues $\{\lambda_n\}$ by:
$\lambda_{n}=\sinh^{2}x_{n}$, which implies $T_{n}=1/\cosh^{2}x_{n}$.
Then, by making the following substitution
\begin{equation}
P(\{x_{n}\},s)~=~\xi(x)~\Psi(\{x_{n}\},s).
\label{BR1}
\end{equation}
with:
\begin{equation}
\xi(x)=\prod_{i<j}|\sinh^{2}x_{j}-\sinh^{2}x_{i}|^{
\frac{\beta}{2}}
\prod_{i}|\sinh 2x_{i}|^{\frac{1}{2}}.
\label{BR2}
\end{equation}
eq.(\ref{DMPK}) can be mapped into a Schr\"{o}dinger
equation (in imaginary time) for the field $\Psi$,

\begin{mathletters}
\begin{eqnarray}
\label{BR3}
-\frac{\partial\Psi}{\partial s}&=&({\cal H}-U)\Psi,
\label{BR3a}\\
{\cal H}&=&-\frac{1}{2\gamma}\sum_{i}\left(\frac{\partial^{2}}
{\partial x_{i}^{2}}+\frac{1}{\sinh^{2}2x_{i}}\right)\nonumber\\
&+&\frac{\beta(\beta-2)}{2\gamma}\sum_{i<j}
\frac{\sinh^{2}2x_{j}+\sinh^{2}2x_{i}}
{(\cosh 2x_{j}-\cosh 2x_{i})^{2}},\label{BR3b}\\
U&=&-\frac{N}{2\gamma}-N(N-1)\frac{\beta}{\gamma}-
N(N-1)(N-2)\frac{\beta^{2}}{6\gamma}.\label{BR3c}
\end{eqnarray}
\end{mathletters}%

By choosing $\beta=2$, the remaining interaction terms among the $x_n$
disappear, the equation can be decoupled and, by imposing ballistic
initial conditions, one ends up with the solution:

\begin{eqnarray}
&&P(\{x_{n}\},s)=C(s)\prod_{i<j}(\sinh^{2}x_{j}-\sinh^{2}x_{i})
\prod_{i}(\sinh 2x_{i})\times\nonumber\\
&&{\rm Det}\left[\int_{0}^{\infty}\!\!dk\,
e^{-\frac{k^{2}s}{4N}}\tanh(\frac{\pi k}{2})k^{2m-1}\,
{\rm P}_{\frac{1}{2}({\rm i}k-1)}(\cosh 2x_{n})\right]
\label{BR4}
\end{eqnarray}
where $P_{\nu}(z)$ are the Legendre functions of the first kind.
 $C(s)$ is a function which depends only on $s$. It can be fixed
by imposing the requirement that $P(\{x_{n}\},s)$
is normalized to unity, and
does not play any role. For this reason we shall
neglect in the following similar multiplicative functions
when writing $P(\{x_{n}\},s)$.

The aim of this letter is to show that the Hamiltonian $\cal H$ defined
above exactly coincides with that of
a particular Calogero-Sutherland (CS) model~\cite{cs}
(for a comprehensive review see
ref.~\cite{op}) with $\sinh$-type interactions
and a potential of type $C_N$ (see
below). This model describes  $N$ particles on a line,
identified by their
coordinates $\{x_i\},~~i=1\cdots N$, interacting with the
potential $1/\sinh^2(x_i-x_j)$.  Its most relevant feature
is that (under particular conditions discussed below, see
eq.(\ref{GM})) it has $N$
commuting integrals of motions, it is completely integrable and its
Hamiltonian can be mapped into the radial part of a
Laplace-Beltrami operator on a
suitable symmetric space.
We shall see below that in our case this Laplace-Beltrami operator is
related in a simple way to the DMPK operator $D$ of eq.(\ref{DMPK}).
This allows to write   the eigenvalue distribution $P(\{x_{n}\},s)$
as a superposition ``zonal spherical functions''  of the
symmetric space.
By using the asymptotic expansion of these functions one can eventually
obtain the explicit expression for  the eigenvalue distribution both in
the metallic and in the insulating regimes.

In the original
formulation of the CS model, the interaction among the particles was
simply pairwise~\cite{cs}. But it was later realized that the
complete integrability of the model had a deep group theoretical
explanation, that the simple pairwise interaction was the signature of
an underlying structure: namely the root
lattice of the Lie algebras $A_N$
and that all the relevant properties (complete integrability, mapping
to a Laplace-Beltrami operator of a suitable symmetric space) still
hold for potentials constructed by means of any root lattice
canonically associated to a simple Lie algebra~\cite{op}.

Let us call ${\cal V}$ the $N$ dimensional space defined by the
coordinates $\{x_i\}$ and $x=(x_1,\cdots x_N)$ a vector in ${\cal V}$.
Let $R=\{\alpha\}$ be a root system in ${\cal V}$, and $R_+$  the
subsystem of positive roots of $R$. Let us denote with $x_\alpha$
the scalar product $(\alpha,x)$.
Then the general form of the CS Hamiltonian is
\begin{equation}
H=-\frac{1}{2}\sum_{i=1}^{N}\frac{\partial^2}{\partial^2x_i}+
\sum_{\alpha\in R_+}\frac{g^2_\alpha}{\sinh^2(x_\alpha)}
\label{CS1}
\end{equation}
 where the couplings $g^2_\alpha$ are the same for equivalent roots,
namely for those roots which are connected with each other by
transformations of the Coxeter group $W$ of the root system.

 The CS models which are relevant for our problem are
those constructed by means of the $C_N$ lattice.
In this case the  root system is $R=\{\pm 2e_i,~\pm e_i \pm e_j,
{}~i\neq j\}$, (where $\{e_i,\cdots e_N\}$ denote a
canonical basis in the space {\bf R}$^n$) and the Coxeter group
$W$ coincides with the product of the permutation group and the group of
transformations which change the sign of the vectors $\{e_i\}$.
The corresponding Hamiltonian is:

\begin{eqnarray}
H&=&-\frac{1}{2}\sum_{i=1}^{N}\frac{\partial^2}{\partial^2x_i}+
\sum_i\frac{g_2^2}{\sinh^2(2x_i)}\nonumber\\
&+&\sum_{i<j}\left(\frac{g_1^2}{\sinh^2(x_i-x_j)}
+\frac{g_1^2}{\sinh^2(x_i+x_j)}\right)~~.
\label{CS3}
\end{eqnarray}

By using  simple identities among hyperbolic functions
eq.(\ref{CS3}) can be rewritten as follows:
\begin{eqnarray}
H&=&-\frac{1}{2}\sum_{i=1}^{N}\frac{\partial^2}{\partial^2x_i}
+\sum_i\frac{g_2^2}{\sinh^2(2x_i)}+c\nonumber\\
&+&2g_1^2\sum_{i<j}
\frac{\sinh^2(2x_i)+\sinh^2(2x_j)}
{\left(\cosh(2x_i)-\cosh(2x_j)\right)^2}
\label{CS4}
\end{eqnarray}
with $c$ an irrelevant constant.
By setting $g_2^2=-1/2$ and $g_1^2=\beta(\beta-2)/4$ we see that
eq.(\ref{CS3}) coincides (apart from the overall factor $1/\gamma$)
with ${\cal H}$ in eq.(\ref{BR3b}).

The CS hamiltonian (\ref{CS3}) can be mapped into the radial part of
a Laplace-Beltrami operator $B$
(see for instance Appendix D of ref.~\cite{op}) of a suitable symmetric
space
\begin{equation}
H=\xi(x)\left[\frac{1}{2}(B+\rho^2)\right]\xi(x)^{-1}
\end{equation}
with
\begin{equation}
B=[\xi(x)]^{-2}\sum_{k=1}^{n}\frac{\partial}{\partial x_k}[\xi(x)]^2
\frac{\partial}{\partial x_k}~~,
\end{equation}
and $\rho$ a constant term which we shall neglect in the following.
At this point it is easy to relate the DMPK  operator of
eq.~(\ref{DMPK}) to the
$B$ operator:
\begin{equation}
D=\frac{1}{2\gamma}~[\xi(x)]^2~ B~ [\xi(x)]^{-2},
\label{DtoB}
\end{equation}
and reobtain  in this way the DMPK equation. This mapping was
discussed in great detail in ref.~\cite{br}. Notice that, as it was
stressed in~\cite{br}, it is not obvious that Sutherland's mapping
should work also for the non-translationally invariant interaction of
eq.(\ref{CS4}). It is exactly the $C_N$ structure underlying
eq.(\ref{CS4}) which allows such a map~\cite{op}.

All the irreducible
symmetric spaces of classical type can be classified
with essentially the same techniques used for the Lie algebras.
They fall into 11 classes labelled by the type of root system and by the
multiplicities of the various roots~\cite{helg}.

A key role in the identification of the DMPK equation as the radial part
of a Laplace-Beltrami operator on a symmetric space
is played by the constants $g^2_\alpha$. In fact
it can be  shown~\cite{op}
that such an identification is possible only if
\begin{equation}
g_\alpha^2=\frac{m_\alpha(m_\alpha-2)}{8}|\alpha|^2
\label{GM}
\end{equation}
where $|\alpha|$ is the length of the root $\alpha$ and $m_\alpha$ its
multiplicity.
In our case we have $m_\alpha=\beta$ for the short roots (those of the
type $\{\pm e_i \pm e_j\}$) and $m_\alpha=1$ for the long roots
(those of the type $\{2 e_i\}$). Remarkably enough these values coincide
with the multiplicities of irreducible symmetric spaces exactly for
those values of $\beta\in\{1,2,4\}$ which are physically relevant.
The identification is as follows:
\begin{eqnarray}
\beta&=&1\hskip0.2cm : \hskip0.5cm Sp(N,\mbox{\bf R})/U(N)\nonumber\\
\beta&=&2\hskip0.2cm : \hskip0.5cm SU(N,N)/S(U(N)\otimes U(N))\nonumber\\
\beta&=&4\hskip0.2cm : \hskip0.5cm SO^{*}(4N)/U(2N)\nonumber
\end{eqnarray}

It is interesting to notice that this same identification was obtained
by H\"uffmann~\cite{huf} by directly looking to the symmetry properties
of the DMPK equations for various values of $\beta$.

According to eq.(\ref{DtoB}) if $\Phi_k(x)$, $x=\{x_1,\cdots,x_N\}$,
$k=\{k_1,\cdots,k_N\}$ is an eigenfunction of
$B$ with eigenvalue $k^2$, then $\xi(x)^2\Phi_k(x)$ will be an
eigenfunction of the DMPK operator with eigenvalue $k^2/(2\gamma)$.
These eigenfunctions  of the $B$  operator  are known in the literature
as ``zonal spherical functions''. In the following we shall use
three important properties of these functions (see~\cite{hc}).

\vskip 0.2cm
\noindent
{\bf 1}] By means of the zonal spherical functions one can define the
analog of the Fourier transform on symmetric spaces:
\begin{equation}
f(x)=\int \bar f(k) \Phi_k(x) \frac{dk}{|c(k)|^2}
\label{ss1}
\end{equation}
(where we have neglected an irrelevant multiplicative constant)
and in the three cases which are of interest for us:
\begin{equation}
|c(k)|^2=|\Delta(k)|^2
\prod_{j}\left\vert\frac{\Gamma\left(
i\frac{k_j}{2}\right)}{\Gamma\left(\frac{1}{2}+
i\frac{k_j}{2}\right)}\right\vert^2
\label{ss2}
\end{equation}
with
\begin{equation}
|\Delta(k)|^2=\prod_{m<j}\left \vert\frac{\Gamma\left(
i\frac{k_m-k_j}{2}\right)\Gamma\left(
i\frac{k_m+k_j}{2}\right)}{\Gamma\left(\frac{\beta}{2}+
i\frac{k_m-k_j}{2}\right)\Gamma\left(\frac{\beta}{2}+
i\frac{k_m+k_j}{2}\right)}\right\vert^2
\label{ss2b}
\end{equation}
where $\Gamma$ denotes the Euler gamma function.

\vskip 0.2cm
\noindent
{\bf 2}] for large values of $x$,
$\Phi_k(x)$ has the following asymptotic behaviour:
\begin{equation}
\Phi_k(x)\sim \frac{1}{\xi(x)} \left(\sum_{r\in W} c(r k) e^{i(r k,x)}
\right)~~,
\label{ss3}
\end{equation}
where $r k$ is the vector obtained acting with $r\in W$ on $k$.
The important feature of eq.(\ref{ss3}) is that it is valid for {\it all
values of k}.

\vskip 0.2cm
\noindent
{\bf 3}]
in the case $\beta=2$ the explicit form of $ \Phi_k(x)$
is known~\cite{op,bk}:
\begin{equation}
\Phi_k(x)=\frac{det\left[ Q_m^j\right]}
{\prod_{i<j}[(k_i^2-k_j^2)(\sinh^2 x_i
-\sinh^2 x_j)]}
\label{ss4}
\end{equation}
where the matrix elements of $Q$ are:
\begin{equation}
Q_m^j={\rm F}\left(\frac{1}{2}(1+ik_m),
\frac{1}{2}(1-ik_m),1;-\sinh^2x_j\right)
\label{ss4b}
\end{equation}
and $F(a,b,c;z)$ is the hypergeometric function.

\vskip 0.2 cm
Eq.s(\ref{DtoB},\ref{ss1}-\ref{ss2b}) allow to write the $s$-evolution of
$P(\{x_n\},s)$  from  given initial conditions (described by the
function $\bar f_{0}(k)$) as follows:
\begin{equation}
P(\{x_n\},s)=[\xi(x)]^2 \int \bar f_0(k)e^{-\frac{k^2}{2\gamma}s}
 \Phi_k(x) \frac{dk}{|c(k)|^2}.
\label{ss5}
\end{equation}
 By inserting the explicit expression of
$|c(k)|^2$ and by using the identity:
\begin{equation}
\left\vert\frac{\Gamma\left(\frac{1}{2}+
i\frac{k}{2}\right)}
{\Gamma\left(
i\frac{k}{2}\right)}\right\vert^2=\frac{k}{2}\tanh\frac{\pi k}{2}
\end{equation}
we end up with the following general expression for $P(\{x_n\},s)$
with ballistic initial conditions (which, due to the normalization
of $\Phi_k(x)$, simply amount to choosing $ \bar f_0(k)=const$):

\begin{equation}
P(\{x_n\},s)=[\xi(x)]^2 \int dk
 e^{-\frac{k^2}{2\gamma}s}
\frac{ \Phi_k(x)}{|\Delta(k)|^2} \prod_{j}k_j\tanh(\frac{\pi k_j}{2})
\label{ss6}
\end{equation}
This is the main result of this letter. Let us try now a few
applications.

First, as a consistency check of the whole approach,
if we insert, in the $\beta=2$ case, the explicit expression
for $\Phi_k(x)$, given in eq.s(\ref{ss4},\ref{ss4b}),
into eq.(\ref{ss6}) and use the identity
\begin{equation}
{\rm P}_{\nu}(z)=F(-\nu,\nu+1,1;(1-z)/2)
\end{equation}
 we exactly
obtain (as expected) the solution, eq.(\ref{BR4}) found by Beenakker and
Rejaei.

Second, if $x$ is large (and in our framework this means $x^2> (2s)/
\gamma$) we may insert  the asymptotic
expansion (\ref{ss3}) into eq.(\ref{ss6}).
 The resulting behaviour of $P(\{x_{n}\},s)$  will depend
on the chosen (metallic or insulating) regime for $k$.
 In both cases,
by setting $\beta=2$ in eq.s (\ref{f1}) and (\ref{f2}) below,
we exactly recover the
results described in~\cite{br}, in these regimes.
Let us look at the two cases  separately.

\vskip 0.2cm

{\it Insulating regime $(k\ll 1)$.}

\noindent
In the $k \to 0$ limit the $\Gamma$ functions in eq.(\ref{ss3})
can be approximated according to:
\begin{equation}
\frac{\Gamma(\frac{\beta}{2}+iy)}{\Gamma(iy)}\sim_{y\to 0}
iy~~,~~\beta\in\{1,2,4\}~.
\end{equation}
Then, by
rewriting both the
product $\prod_{i<j}(k^2_i-k^2_j)$, and the sum over the exponentials in
(\ref{ss3}) as  determinants,
the  integration over $k$ becomes straightforward and gives:

\begin{eqnarray}
P(\{x_{n}\},s)&=&\prod_{i<j}\left\vert\sinh^{2}x_{j}-\sinh^{2
}x_{i}\right\vert^{\frac{\beta}{2}}\left[
(x_{j}^{2}-x_{i}^{2})\right]\nonumber\\
&\times &\prod_{i}\left[\exp(-x_{i}^{2}\gamma/(2s))x_{i}(\sinh
2x_{i})^{1/2}\right].
\label{f1}
\end{eqnarray}

 Ordering the
$x_{n}$'s from small to large and using the fact that in this regime
 $1\ll x_{1}\ll x_{2}\ll\cdots\ll x_{N}$ we can
approximate the eigenvalue distribution as follows:
\begin{equation}
P(\{x_{n}\},s)=\prod_{i=1}^{N}
\exp\left[-(\gamma/(2s))(x_{i}-\bar{x}_{i})^{2}\right].
\label{IR2}
\end{equation}
where $\bar{x}_{n}=\frac{s}{\gamma}(1+\beta(n-1))$,
in agreement with the result obtained by Pichard,\cite{pic} by directly
solving the DMPK equation in this regime.

\vskip 0.2cm

{\it Metallic regime $(k\gg 1)$.}

\noindent
In this case one must use
the asymptotic expansion:
\begin{equation}
\frac{\Gamma(\frac{\beta}{2}+iy)}
{\Gamma(iy)}\sim_{y\to\infty} |y|^{\frac{\beta}{2}}~e^
{\frac{i\pi\beta}{4}}~,~~\beta\in\{1,2,4\}~.
\end{equation}
The integration over $k$ is less simple in this case and,
(to be consistent with the regime of validity of eq.(\ref{ss3}))
in the resulting expression
only the highest powers of $(x\sqrt{\frac{\gamma}{2s}})$
must be taken into account.
We find:
\begin{eqnarray}
P(\{x_{n}\},s)&=&\prod_{i<j}\left\vert\sinh^{2}x_{j}-
\sinh^{2}x_{i}\right\vert^{\frac{\beta}{2}}\left\vert
x_{j}^{2}-x_{i}^{2}\right\vert^{\frac{\beta}{2}}\nonumber\\
&\times&\prod_{i}\left[\exp(-x_{i}^{2}\gamma/(2s))(x_{i}\sinh
2x_{i})^{1/2}\right].
\label{f2}
\end{eqnarray}
In agreement with the exact result of ~\cite{br} for $\beta=2$ and with
the $\beta$ dependence found by Chalker
and Mac\^{e}do~\cite{cm2} through a direct integration of the DMPK
equation.

According to  ref.~\cite{op} and ref.~\cite{hc}, the regime of validity
of eq.s(\ref{f1}) and (\ref{f2})  is  $x^2> (2s)/\gamma$, which gives in
in the large $N$ limit $x^2>(2s)/(\beta N)$.
Notice however, as a concluding remark,
that eq.(\ref{ss3}) is only the first term of a series
which converges absolutely to  $\Phi_k(x)$ for all values of $k$ (see
sect. 8 of ref.~\cite{op} and ref.~\cite{hc}).
The coefficients of
this series can be constructed recursively, thus allowing
to study the behaviour of eq.(\ref{ss6}) even for values of $x$ smaller
than the above mentioned threshold.

\end{document}